\documentclass[twocolumn,prb,showpacs]{revtex4}

\usepackage{graphicx}

\begin{document}

\newcommand*{\cm}{cm$^{-1}$}
\newcommand{\comment}[1]{}

\title{Infrared and optical properties of pure and cobalt-doped
LuNi$_2$B$_2$C.}

\author{M.~Windt}
\altaffiliation[Present address: ]{II. Physikalisches Intitut, Universit\"at zu of K\"oln,
50937 k\"oln, Germany}

\author{J. J. McGuire}

\author{T. R\~o\~om}
\altaffiliation[Present address: ]{National Institute of Chemical Physics and Biophysics,
12618 Tallinn, Estonia}

\author{A. Pronin}
\altaffiliation[Present address: ]{Institute of General Physics,
Russian Academy of Sciences, 119991, Moscow, Russia}

\author{T. Timusk}
\altaffiliation{The Canadian Institute of Advanced Research}
\email{timusk@mcmaster.ca}

\affiliation{Department of Physics and
Astronomy, McMaster University, Hamilton ON Canada, L8S 4M1}

\author{I.R. Fisher}
\author{P.C.~Canfield}

\affiliation{Ames Laboratory,
 Department of Physics and Astronomy,Iowa State University, Ames, Iowa 50011}

\begin{abstract}
We present optical conductivity data for
Lu(Ni$_{1-x}$Co$_x$)$_2$B$_2$C over a wide range of frequencies
and temperatures for x=0 and x=0.09. Both materials show evidence
of being good Drude metals with the infrared data in reasonable
agreement with dc resistivity measurements at low frequencies. An
absorption threshold is seen at approximately 700 \cm. In the
cobalt-doped material we see a superconducting gap in the
conductivity spectrum with an absorption onset at $24 \pm 2$ \cm\
= 3.9$\pm 0.4 k_BT_c$ suggestive of weak to moderately strong
coupling. The pure material is in the clean limit and no gap can
be seen. We discuss the data in terms of the electron-phonon
interaction and find that it can be fit below 600 \cm\ with a
plasma frequency of 3.3 eV and an electron-phonon coupling
constant $\lambda_{tr}=0.33$ using an $\alpha^{2}F(\omega)$
spectrum fit to the resistivity.
\end{abstract}


\maketitle

\section{Introduction}

The recent discovery of superconductivity at 39 K in MgB$_2$ has
renewed interest in intermetallic superconductors containing
boron. There have been suggestions that their high transition
temperatures may result from an exotic mechanism and not from a
conventional s-wave BCS phonon process. The borocarbide family of
superconductors LNi$_2$B$_2$C, where L=(Y,Lu,Tm,Er,Ho,and Dy)
shows a number of physical properties that suggest that they may
be a testing ground for these ideas.\cite{canfield98} For example,
recent measurements of heat capacity and microwave surface
impedance in the vortex state suggest that low energy
quasiparticles may have d-wave dispersion.\cite{wang98,izawa01}
The critical field temperature variation has an upward curvature
that is difficult to explain within the BCS
theory.\cite{wang98,manolo99} Other suggestions of abnormal
behavior include residual absorption in the gap seen by Raman
spectroscopy\cite{yang99a,yang99} and the sensitivity of the
transition temperature to non-magnetic
impurities.\cite{schmidt94,cheon98} On the other hand, there is a
lot of the evidence that supports a conventional s-wave mechanism
of superconductivity for these materials.  This includes good
agreement between transport and critical field properties based on
Eliashberg theory\cite{shulga98,gonnelli00} and s-wave-like
tunneling spectra.\cite{suderow01}

Within the s-wave, electron-phonon picture, there are two routes
to high transition temperatures: a large coupling constant
$\lambda$ or the coupling to very high frequency modes with modest
coupling.  It was suggested early on by Pickett \textit{et al.} that
unusual electronic structure seen in the borocarbides at the Fermi
surface would lead to strong coupling to certain phonons.\cite{pickett94}
On the other hand more recent data based on specific heat by Michor
\textit{et al.} are consistent with a $\lambda$ of the order of
unity.\cite{michor95} Gonnelli \textit{et al.} were able to fit the
temperature dependence of the dc resistivity of YNi$_2$B$_2$C with
a phonon spectrum from inelastic neutron scattering of Gompf
\textit{et al.}\cite{gompf97} with a $\lambda_{tr}$ of only
0.57,\cite{gonnelli00} where $\lambda_{tr}$ is just $\lambda$
modified by a factor that depends of the scattering
angle.\cite{allen71} Gompf \textit{et al.} found evidence of soft
mode behavior in several modes but not the high frequency mode
originally singled out by Mattheiss\cite{mattheiss94} to couple
strongly to the electronic system.

The infrared response of superconductors can be used to sort out
some of these issues. For example, there are striking differences
between exotic and conventional superconductors in the effect on the
infrared response of doping with impurities that limit the scattering
time of the quasiparticles. Doped d-wave superconductors do not develop
a sharp gap signature at $\omega=2\Delta$ as they approach the
dirty limit where the scattering rate $1/\tau > 2\Delta$ as
described by Mattis and Bardeen.\cite{mattis58} Instead of a
region of perfect reflectance followed by a sharp onset of
absorption at $2\Delta$, as seen in conventional materials, d-wave
matreials show a low-frequency Drude-like absorption from defects
surrounded by regions of normal
material.\cite{basov98,schachinger01} For this test it is
necessary to measure the reflectance accurately in the gap region
of doped samples in the region of $\hbar\omega=2\Delta$.

It has been suggested\cite{schmidt94} that the sensitivity of $T_c$
to non-magnetic defects in the borocarbides is due to changes in
$N(0)$, the density of states at the Fermi level, that are related to
a peak in the density of states near the Fermi energy $E_F$ seen in
band structure calculations.\cite{mattheiss94,pickett94}
One effect is the valence effect, where, in the rigid band picture,
doping with holes or electrons will sweep $E_F$ across this peak and
change $N(0)$. Another is where doping with non-magnetic impurities
broadens the peak which leads to the reduction of $N(0)$ which in
turn reduces $T_c$. One of the advantages of infrared spectroscopy
over dc transport measurements is that it allows a separate determination
of the scattering rate $1/\tau$ and the free carrier plasma frequency
$\omega_p$ which is related to $N(0)$. Any dramatic changes to the plasma
frequency with doping would be seen as a change in the Drude spectral
weight with doping. There will also be changes to the slopes of the
resistivity curves with temperature which, in accord with
Matthiessen's rule, should not change with doping if it affects
the carrier life time but not the density of states.

Infrared spectroscopy can also be used to map out the spectrum of
excitations that couple to the electrons. It was shown that BCS
superconductors have peaks in the second derivative of the
absorption spectrum ($A=1-R$, where $R$ is the
reflectance)\cite{joyce70,farnworth76} at the frequencies of
longitudinal and transverse phonons, thus confirming the mechanism
of electron-phonon interaction in these materials. For this test
one has to be able to measure the second derivative of the
reflectance in the region of the relevant excitations. To get the
good signal-to-noise ratio necessary for such experiments large
single crystals are needed.

Several studies of the optical conductivity of the borocarbides
have been published.\cite{widder95,bommeli97,kim00} Widder
\textit{et al.}\cite{widder95} reported on reflectivity and electron
loss spectroscopy on ceramic samples of LuNi$_2$B$_2$C at room
temperature for frequencies up to 50 eV. They find an overall
metallic response with a plasma frequency of 4.25 eV and an
electron-phonon coupling parameter $\lambda_{tr}=1.2$ . Bommeli
\textit{et al.} measured both LuNi$_2$B$_2$C and YNi$_2$B$_2$C over a
wide range of temperatures and frequencies\cite{bommeli97} reporting a
strong absorption in the pure LuNi$_2$B$_2$C material with an
onset of 100 \cm, well below any significant phonon density of
states. They also claim to see a dirty limit superconducting gap
in the undoped material while, according to transport
measurements,\cite{cheon98} the undoped materials should be in the
clean limit. In a recent paper Kim \textit{et al.} show measurements
at room temperature in LuNi$_2$B$_2$C\cite{kim00} supplementing
infrared reflectance with ellipsometry measurements at higher
frequency. They also studied the effect of annealing on their
polished samples. They reported a screened plasma frequency of
3.76 eV for an annealed sample and 3.01 eV for an unannealed
sample. Both samples were polished. They find a Drude-like
frequency-independent scattering rate below 0.3 eV but a gradual
rise in scattering above this frequency signaling non-Drude
behaviour in the midinfrared.

Cheon \textit{et al.} reported on dc transport in a series of cobalt
doped LuNi$_2$B$_2$C single crystals.\cite{cheon98} They found a
dc resistivity that was linear in temperature with an intercept on
the temperature axis at 35 K at $T=\theta_D/10$.\cite{michor95}
The slope of the resistivity increased only slightly with cobalt
doping, providing evidence that any changes in the density of
states at the Fermi surface with cobalt doping were small.  We
decided to investigate the issue of the superconducting gap in
clean \textit{vs.} dirty samples in the borocarbides using samples
from the Ames group, the same source as those used in the Bomelli
\textit{et al.} work.

\section{Experiment}

We used flux grown single crystals in this work characterized
extensively by dc resistivity and magnetic
susceptibility.\cite{cho95,cheon98} The reflectance measurements
were done on natural growth faces normal to the c-axis. To correct
for irregularities of the surface, which were mainly vertical
steps, the sample reflectance was referenced to spectra where the
sample was coated with a gold layer, evaporated
\textit{in situ}.\cite{homes93a} The as-grown crystals have tiny
droplets of Ni$_2$B on them.  The visible ones occupy a small
fraction of the surface area and our gold evaporation technique
tends to suppress curved surfaces in favour of flat ones. Metallic
droplets that are much smaller than the wavelength can give rise to
an absorption band in the region of $\omega_p/\sqrt{3}$.

Most of the measurements were carried
out in a cold-finger flow cryostat with a minimum sample temperature
of 10 K. A home made rapid-scan interferometer was used in the
far-infrared with He-cooled bolometer detectors below 700 \cm\ and an
MCT detector up to 8000 \cm. High frequency measurements were
done with a grating spectrometer, and in the superconducting
state, a polarizing interferometer and an immersion
dewar with a He-3 bolometer were used between 5 and 60 \cm.

\begin{figure}
 \resizebox{!}{7cm}{\includegraphics{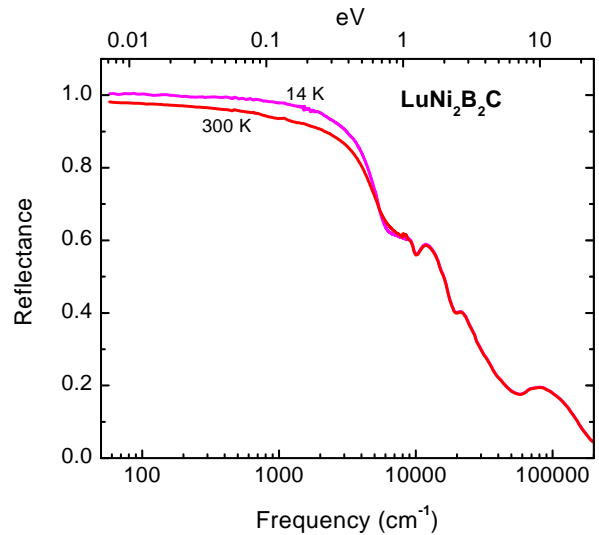}}
 \caption{\label{fig1}The reflectance of LuNi$_2$B$_2$C at two temperatures.
 The data above 5 eV is from Widder \textit{et al.}\cite{widder95}}
\end{figure}

Figure \ref{fig1} shows the reflectance of a crystal of pure LuNi$_2$B$_2$C
over a wide range of frequencies at ambient temperature and at 14
K. Above 5 eV we  have merged our data with the EELS data of
Widder \textit{et al.}\cite{widder95} The data showed little
temperature dependence at 8000 \cm\ and we made only room
temperature measurements above this frequency. At low frequency
the reflectance in the different spectral regions, measured with
different spectrometers and detectors, generally agreed to within
0.5 \% in the region of overlap, whereas at higher frequencies the
mismatch could be as much as several percent.

\begin{figure}
 \resizebox{!}{7cm}{\includegraphics{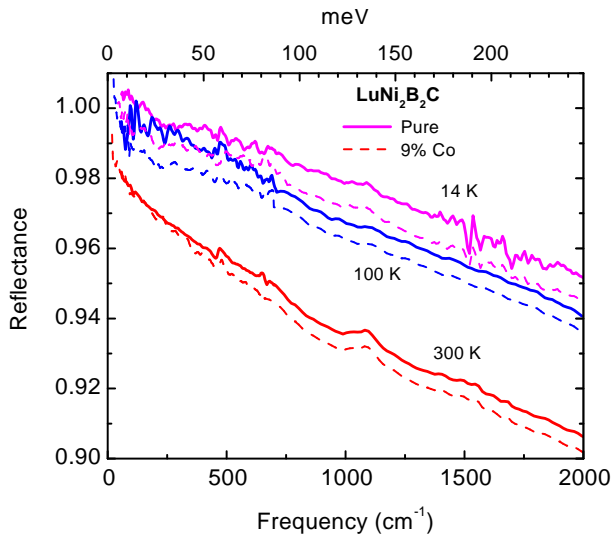}}
 \caption{\label{fig2}Low-frequency reflectance of LuNi$_2$B$_2$C at three temperatures. The solid curves
 are for the undoped sample and the dashed ones for a sample with 9 \% cobalt doping. The fine
 structure is noise.}
\end{figure}

Figure \ref{fig2} shows the low frequency region on an expanded scale at
three temperatures. The solid lines refer to the pure sample of
LuNi$_2$B$_2$C and the dashed curves to the sample with 9 \% Co
doping.  The fine structure on the curves is noise.

A comparison of our measured reflectance with data from other
groups shows in general reasonably good agreement with most of the
previous work but with some notable discrepancies.  First, our
reflectances are in general higher than what has been reported by
previous investigators. Part of this difference we attribute to
our use of as-grown single crystal surfaces since polishing will
introduce surface damage leading to additional absorption
particularly at higher frequencies. Kim \textit{et al.} find that by
annealing the polished crystals a substantial increase in
reflectance can be obtained.\cite{kim00} We have found, in the
past, that to obtain good reflectance spectra of metals in the
ultraviolet it is necessary to electropolish the
samples.\cite{philipp61} The reflectances of our as-grown single
crystal surfaces are much higher than the polished samples leading
to conductivities in  the interband region that are as much as
three times higher than those of the polished samples.

At 800 cm$^{-1}$ at room temperature, we measure a reflectance of
94 \% whereas both Kim \textit{et al.} and Bommeli \textit{et al.}
report $\approx$91 \% in polished samples. We fail to see the
strong absorption seen by Bommeli \textit{et al.} at low temperature
with an onset at 50 \cm\ where the reflectance falls rapidly from
99 \% at 50 to 93 \% at 150 \cm. In all our samples, doped and
undoped, the low temperature reflectance below 700 \cm\ was above
98 \%, as expected for a good metal.

\begin{figure}
 \resizebox{!}{6cm}{\includegraphics{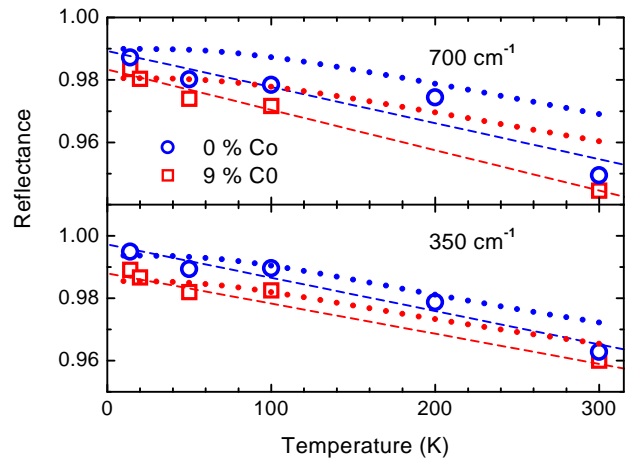}}
 \caption{\label{fig3}The temperature dependence of the reflectance at two frequencies. The circles
 are data for the pure sample, the squares for the cobalt-doped sample. The dashed lines
 are least squares fits to the experimental points. The small points are for
 a theoretical model based on the interaction with a mode at 170 \cm\ with
 parameters fit to dc transport and an assumed plasma frequency of 3.3 eV (top points
 pure and bottom doped samples). }
\end{figure}

\begin{figure}
 \resizebox{!}{7cm}{\includegraphics{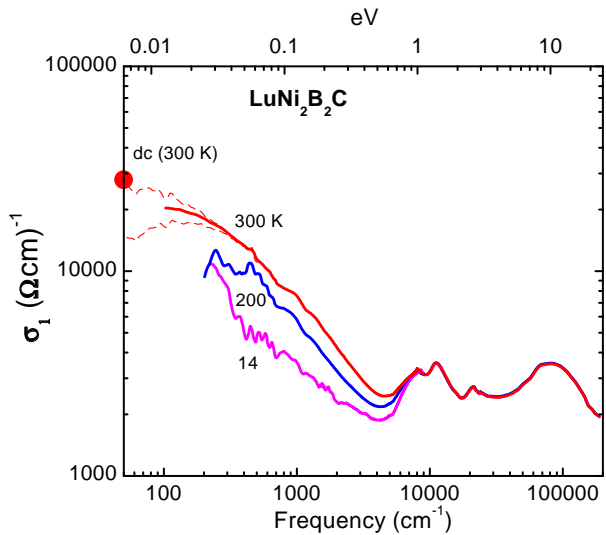}}
 \caption{\label{fig4}The optical conductivity of pure LuNi$_2$B$_2$C at three
 temperatures. The dashed lines show estimated error limits based on
 a 0.5 \% uncertainty of the 100 \% line.  The circular symbol on the
 vertical axis is the dc conductivity at room temperature. The
 decreasing conductivity as the temperature is lowered is due to the narrowing
 of the Drude peak.}
\end{figure}

The overall high reflectance of these single crystals places
demands an an accurate calibration of the absolute value of the
reflectance. We calibrated our system by using polished stainless
steel and a lead indium alloy as primary standards. We started by
measuring the dc resistivity of these reference samples and used a
Drude model to calculate the absolute value of the reflectance.
These calibrations were used to determine the reflectance of our
evaporated gold coating which was our secondary standard. Runs
were rejected where the reflectance changed by more than 0.5 \%
between the beginning and the end of the run. As another check we
compared our reflectances in the low frequency region with the
Drude model based on dc resistivity measurements of crystals
prepared in the same way as shown in Fig. \ref{fig3} and Fig. \ref{fig4}.

Figure \ref{fig3} gives a picture of the reproducibility of our
measurements.  We have plotted the reflectance at 350 cm$^{-1}$
and at 700 cm$^{-1}$  as a function of temperature for our two
samples. In this spectral region $\omega\tau>1$ and
$R=1-2/(\omega_p\tau)$. As the discussion below shows, we find a
plasma frequency of $\omega_p=3.3$ eV. With this value we can
relate our reflectance measurements directly to dc resistivity
($\rho=4\pi/(\omega_p^2\tau)$). The predicted reflectances from
the dc resistivity are shown as small symbols in Fig. \ref{fig3} for the
two samples. We see generally excellent agreement in the
temperature dependence as well as the difference in the two sample
curves due to the doping induced elastic scattering. The dashed
curves are a least squares fit of a straight line to the
experiments. The root mean square deviation from the straight line
is less than 0.15 \% for all the curves. However the high
frequency curves fall systematically below the dc transport
prediction.  This is due to the rising 0.15 eV mid-infrared band
that has a threshold in this spectral region.

To obtain the optical conductivity we performed Kramers-Kronig
analysis of the reflectance spectra. For this it is necessary to
extend the data at low and high frequencies.  At low frequencies
we assumed a Drude model where the parameters were fit to the
lowest measured frequency,  60 cm$^{-1}$ for the undoped sample
and 20 cm$^{-1}$ for the cobalt doped sample. At high frequency,
from 5 eV to 35 eV, we used the data of Widder
\textit{et al.}\cite{widder95} and beyond that a power law $R \propto
\omega^{-1.5}$ to $\omega=100$ eV and $R \propto \omega^{-4}$
above 100 eV.

Figure \ref{fig4} shows the optical conductivity for the undoped sample at
300, 200 and 14 K for the whole measured range.  The dashed curves
are estimated errors based on a 0.5 \% error in the absolute value
of reflectance. The solid point is the dc resistivity measurement
at room temperature of Cheon \textit{et al.}\cite{cheon98} Our
optical data match the dc data within our experimental error. The
error of the dc data is $\pm 10 $\% and mainly due to
uncertainties in contact geometry. As the temperature is lowered
the conductivity drops in the midinfrared and rises in the far
infrared. We are unable to show far infrared conductivities below
200 \cm\ because the reflectance becomes too high for accurate
Kramers-Kronig analysis.

\begin{figure}
 \resizebox{!}{7cm}{\includegraphics{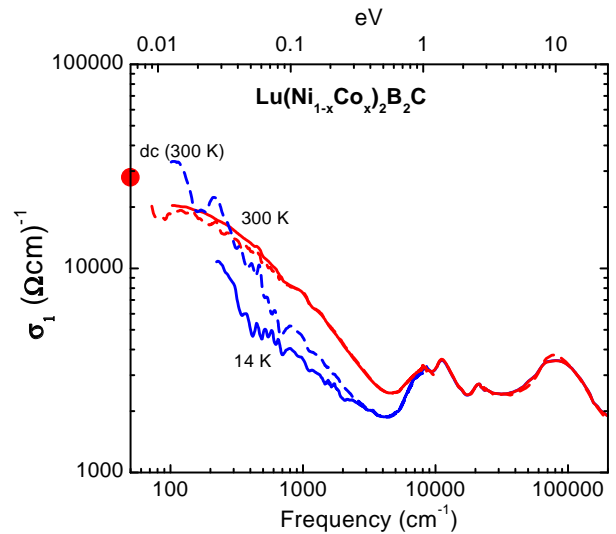}}
 \caption{\label{fig5}The conductivity of the 9 \% cobalt-doped sample (dashed lines) and the
 undoped sample, (solid lines). The solid ball on the vertical axis is the
 dc conductivity. While there is little difference at room temperature, the Drude
 peak of the doped
 sample is substantially broader at low temperature.}
\end{figure}

Figure \ref{fig5} shows the conductivity of the doped sample at two
temperatures, along with the pure sample, shown with solid lines.
It can be seen clearly that while the doping has little effect at
room temperature, at low temperature the Drude peak is broader as
expected from the additional scattering.


\begin{figure}
 \resizebox{!}{7cm}{\includegraphics{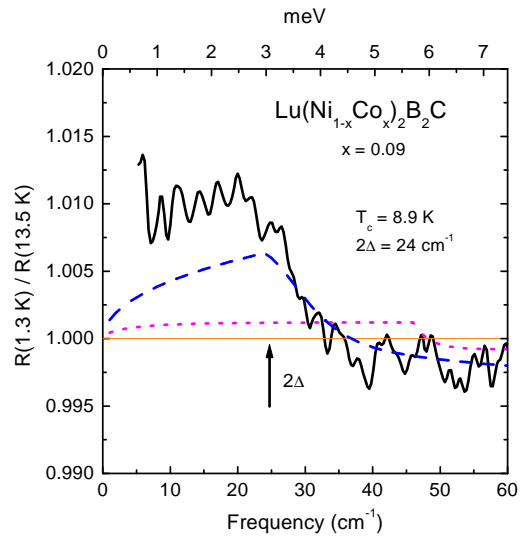}}
 \caption{\label{fig6}The reflectance in the superconducting state at
 1.3 K divided by the normal state reflectance at 13.5 K. The
 superconducting gap can be seen as an onset of absorption in the
 normal state.  A BCS theoretical curve for this sample is shown
 as the curve with the long dashes.  The dotted curve is a
 calculation for the undoped sample. The undoped sample is in the
 clean limit and the gap signature is predicted to be weak.}
\end{figure}

The measurements with the polarizing spectrometer in the
superconducting state for the 9 \% doped sample are shown in Fig.
\ref{fig6}. To increase the sensitivity we only show a reflectance ratio
between the superconducting state and normal state shown as the
solid curve in the figure.  The dirty limit BCS superconducting
gap signature can be seen clearly as a drop in the ratio at 22
\cm. The dashed curve shows the calculated reflectance ratio
based on dc conductivity of the 9 \% doped sample. We estimate a
gap value $2\Delta=24 \pm 2$ \cm.

The disagreement in our data of the peak amplitude in the observed
reflectance ratio with a BCS calculation is 0.5 \%. This
difference is larger than we expect from such a measurement where
the sample is not moved and the two measurements are done within
minutes of one another. We have measured a lead indium alloy in
the same geometry without observing a discrepancy between the dc
transport prediction and the reflectance ratio. The calculation
assumes the superconductor to be a perfect reflector in the gap
region. Any residual absorption in the superconducting state due
to strong gap anisotropy or a second phase on the sample surface
would reduce the absorption below the calculated value and have
the opposite effect to what we observe. An overall decrease in
detector sensitivity with increasing temperature may be partially
responsible for the discrepancy but it would have the effect of
uniformly shifting the ratio upward causing the data above the gap
value to exceed unity. We do not observe such an effect here
although we cannot rule out a 0.2 \% contribution from this cause.
Since we have used dc resistivity to estimate the strength of the
normal state absorption any discrepancy between the resistivity of
our sample and the bulk measurement could explain the discrepancy.
However the agreement at higher frequencies shown in Fig. \ref{fig3} rules
out any large discrepancies.

The model predicts a sharp change of slope at $2\Delta$ whereas
our experiments show what might be two steps, one at 22 cm-1 and
another at 26 cm-1, which may be due to gap anisotropy as proposed
by Yang \textit{et al.}\cite{yang99a} based on Raman spectroscopy
done in different scattering geometries. However scattering due to
the Co dopants would be expected to remove any such anisotropy.

The curve with the long dashes shows the calculated spectrum for
the pure sample. The expected gap signature is weak since the
sample approaches the clean limit with $1/\tau=17$ \cm,
determined from the dc conductivity with $\omega_p=3.3$ eV and
$2\Delta=46$ \cm\ from Raman spectroscopy. We measured the pure
sample reflectance ratio but failed to see any sign of a
superconducting gap in the reflectance ratio of our samples within
the estimated noise level of $\pm$ 0.002 in the reflectance ratio.

\section{Results}

There are two popular ways of analyzing the conductivity if it
does not follow the standard Drude form.  The first, the two
component method, is to fit the conductivity to a Drude peak
centered at zero frequency and add a second component consisting
of a set of oscillators in the mid-infrared to allow for any
additional non-Drude absorption. The second method, the extended
Drude model, introduced by Allen and Mikkelsen,\cite{allen77}
assumes that the Drude form applies throughout the infrared but
that the Drude scattering rate increases with frequency, making,
in effect, the Drude peak broader as the frequency increases. The
first method accounts well for any parallel channels of
conductivity such as optic phonons or low-lying interband
transitions. The second method is needed when the interaction of
the free carriers is strong and a Holstein sideband is formed at
low temperatures. At high temperatures this sideband merges with
the Drude absorption forming an approximate overall Drude-like
band. In the absence of a detailed theory of the optical
conductivity both methods have value in the analysis of the data
if one keeps open the possibility that some of the mid-infrared
oscillators have no real physical meaning or that the increase in
the scattering rate seen in the one-component analysis may be due
to interband processes and not to self energy effects on the free
carriers.

\begin{figure}
 \resizebox{!}{7cm}{\includegraphics{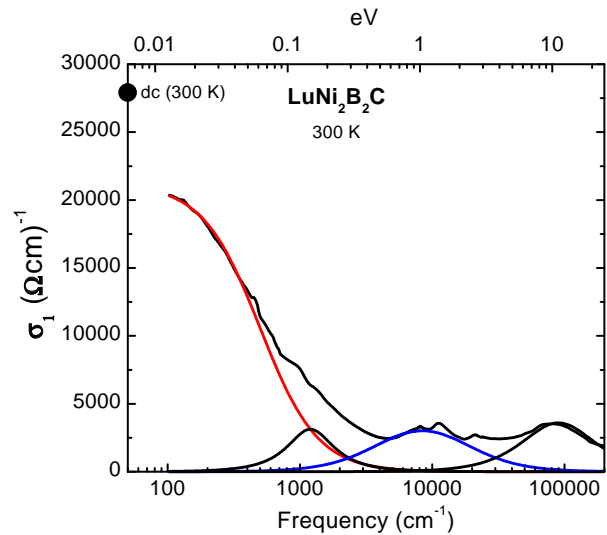}}
 \caption{\label{fig7}Oscillator fits to the reflectance (solid curve). In addition
 to the Drude term at low frequency, three oscillators are needed to
 describe the conductivity at higher frequencies.}
\end{figure}

The two component picture is illustrated in Fig. \ref{fig7} where we show
an oscillator fit to the conductivity for the undoped sample at
room temperature. For the undoped sample at 300 K the following
oscillators were used, (center frequency, width, and plasma
frequency, in \cm): 0, 500, 25200; 1200, 1200, 15000; 8410,
20000, 60000; 90000, 150000, 180000. It is clear that the
simple Drude form does not fit at high frequency and additional
Lorentz oscillators are needed to fit the data. The two high
frequency oscillators can be identified as due to interband
transitions\cite{widder95} since band structure calculations
indicate a gap in the density of states just below the Fermi level
rising to a strong peak about 2 eV below the Fermi level.
Combining this hole density of states with the sharp peak at the
Fermi level would predict an optical conductivity that rises
gradually from low frequencies to form the plateau we see in the 1
to 10 eV region. It is difficult, in the absence of a detailed
joint density of states, to say if there is a low frequency
threshold to the optical conductivity but since the dispersion
curves show several band crossings very close to the Fermi level,
one cannot rule out an interband conductivity that extends to zero
frequency.

The midinfrared oscillator at 0.15 eV is more difficult to assign.
Widder \textit{et al.}\cite{widder95} have assigned the spectral
weight in this region to the frequency dependent scattering rate
of phonons, but as we will show below, it is difficult to get a
consistent picture of the data in this spectral region with a
phonon spectrum that is confined to the known phonon density of
states. Fig. \ref{fig9} (to be discussed below) shows that there is an
absorption threshold at 700 \cm\ at low temperature which
suggesting that interband transitions may set in at this
frequency. However the onset frequency of the threshold is
strongly temperature dependent, moving to 100 \cm\ at 300 K. This
temperature dependence is difficult to explain in terms of a
simple Fermi function expected for interband transitions where a
broadening of only $k_BT\approx 200$ \cm\ is expected.

The spectral weight under the 0.15 eV oscillator could also be the
result of a multiphonon side-band of the Drude peak.  From its
spectral weight we estimate an additional mass-enhancement factor
of 1.35 and a total free carrier plasma frequency of 3.64 as
opposed to 3.12 eV if the 0.15 eV oscillator is assigned to
interband transitions.  A complete Eliashberg calculation would be
needed to see if this model will give a consistent picture of the
gap ratio, the coupling constant $\lambda$ and the transition
temperature $T_c$.

There are several other ways of estimating the Drude weight from
our data. The Drude component in Fig. \ref{fig7} has a plasma frequency of
3.14 eV for the undoped sample. If we scale it to agree with the
dc resistivity we get 3.6 eV. Another method uses the temperature
dependence of the reflectance shown in Fig. \ref{fig3} and compares it to
the temperature dependence of the dc resistivity. The derivative
of the dc resistivity is given by $d\rho/dT= {4\pi \over
\omega_p^2}d\gamma/dT$ where $\gamma$ is the scattering rate. On
the other hand the reflectance in the relaxation region, where
$\gamma < \omega$ is given by $1-R=2\gamma/\omega_p$, where $R$ is
the reflectance. This equation can be differentiated with respect
to $T$ and the relaxation rate derivative $d\gamma/dT$ can be
eliminated between these two equations to give
$\omega_p=2\pi(1-R)'/\rho'$ where the primes denote temperature
derivatives. This method is not subject to errors in absolute
reflectance as it uses the temperature variation as input. Using
this method with the fitted curves in Fig. \ref{fig3}. along with the
resistivity data of Cheon \textit{et al.}\cite{cheon98} we obtain an
average plasma frequency of 3.4 $\pm$ 0.15 eV directly from the
reflectance without using data from Kramers-Kronig analysis. As an
average of the two methods we adopt the plasma frequency of our
pure sample of 3.3 $\pm$ 0.2 eV.

Our results are lower than the 4.25 eV value of Widder \textit{et al.}
who included the 0.15 eV peak in the Drude spectral weight.
Bommeli \textit{et al.} find a two-component low-frequency spectrum
with a Drude weight of 0.73 eV and an additional mid infrared band
with a total oscillator strength of 4 eV by including the broad
absorption at 500 \cm. We have found no evidence for this low
frequency component. Kim \textit{et al.} using polished and annealed
samples find $\omega_p = 3.76$ eV for the screened plasma
frequency determined from the zero crossing of $\epsilon_1$. The
zero crossing of $\epsilon_1$, obtained from Kramers-Kronig
analysis of our reflectance occurs at the same frequency.  However
to get the unscreened plasma frequency $\omega_p$ a correction has
to be made for the very high frequency dielectric constant which
is not known accurately.



The frequency-dependent scattering rate formalism starts with the
extended Drude formula:
$$
\sigma(\omega,T)= \nonumber \\
{1 \over {4\pi}} {{\omega_p^2} \over
{1/\tau(\omega,T)-i\omega[1+\lambda_{tr}(\omega,T)]}},
$$
where $1/\tau(\omega,T)$ is the frequency dependent scattering
rate and $1+\lambda_{tr}(\omega,T)=m^*/m$ is the mass
renormalization factor. These can be calculated from the real and
imaginary parts of the optical conductivity:

$$
1/\tau(\omega)={\omega_p^2 \over {4\pi}} Re({1 \over
\sigma(\omega)}),
$$
and
$$
1+\lambda_{tr}(\omega)=-{\omega_p^2 \over {4\pi}} {1 \over \omega}
Im({1 \over \sigma(\omega)}).
$$

\begin{figure}
 \resizebox{!}{6cm}{\includegraphics{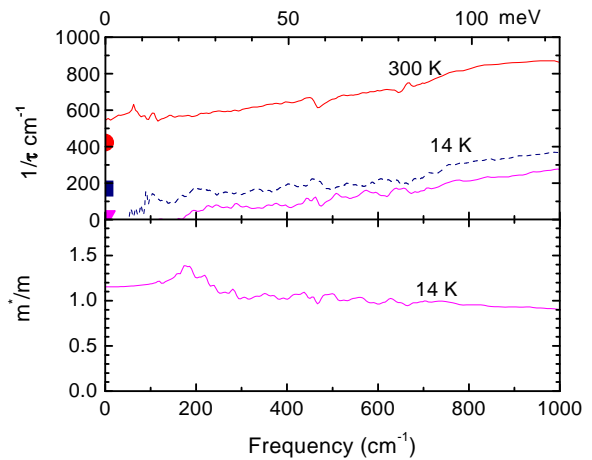}}
 \caption{\label{fig8}Frequency dependent scattering rate, top panel, and the
 effective mass, lower panel, obtained from the extended Drude
 formula, assuming a plasma frequency of 3.3 eV. The dashed line is
 for the cobalt-doped sample. The symbols on the axis are dc values of
 the scattering rate, corresponding to the optical conductivity curves.}
\end{figure}

\begin{figure}
 \resizebox{!}{7cm}{\includegraphics{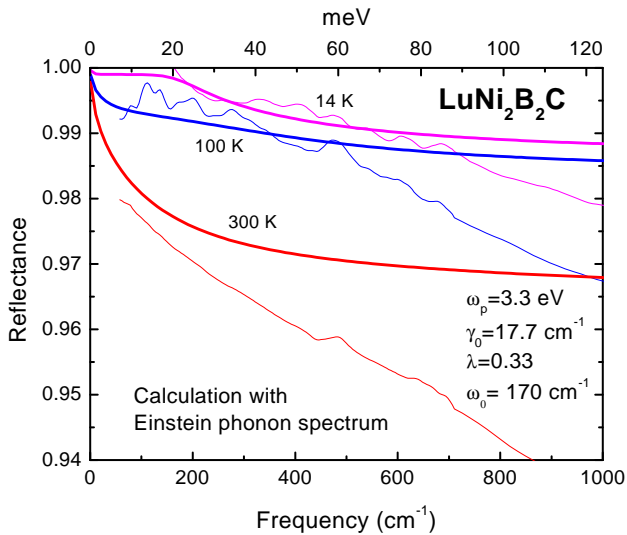}}
 \caption{\label{fig9}Calculated reflectance based on the electron-phonon interaction
 with a coupling constant $\lambda_{tr}$ obtained from the dc resistivity assuming
 an Einstein spectrum of a single phonon at 170 \cm. The agreement with the measured
 reflectance (thin curves) is good at low frequency but strong additional absorption sets
 in at 700 \cm\ at 14 K}
\end{figure}

The frequency dependent scattering rate is shown in Fig. \ref{fig8} along
with the mass renormalization. Both are based on an assumed plasma
frequency of 3.3 eV for the  free carriers and a high frequency
dielectric constant of 1.0. The choice of plasma frequency affects
the overall scale of the curves. The choice of high frequency
dielectric constant has little effect in this frequency region
where $\epsilon_1$ and $\epsilon_2$ are large. We see a monotonic
increase in scattering rate with frequency. The solid curves are
for the pure sample and the dashed one for the doped one.  The
symbols on the vertical axis are the values from the dc
conductivity assuming a plasma frequency of 3.3 eV. \comment{The
dashed curve is the calculated scattering rate based on the phonon
model.}

The lower panel of Fig. \ref{fig8} shows $m^*/m=1+\lambda_{tr}$ for the
pure sample at low temperature, which is, within a factor of
$\omega$, the imaginary part of $1/\tau$.  One can read off the
calculated electron-phonon coupling constant for dc transport, the
zero-frequency value, $\lambda_{tr}(0)$ to be 0.15. The value
obtained more directly from dc transport is 0.34, based on a
plasma frequency of 3.3 eV and the assumption that the transport
curves have reached their high frequency linear limit at 300 K
when $1/\tau=2\pi k_BT$. A strong contribution from high frequency
phonons would raise the dc value of $d\rho/dT$ and lead to an even
larger $\lambda_{tr}$. We do not feel that discrepancy between the
transport value and the optical value is significant since the
curves in Fig. \ref{fig7} have large uncertainties due to the high
reflectance of the undoped sample at low frequencies.

It is important to stress the difference between the curves in
Fig. \ref{fig8} and the corresponding data for the
cuprates.\cite{puchkov96d} First, the overall scale of scattering
in the cuprates is an order of magnitude higher to the point where
the scattering rate is larger than the frequency, leading to a
breakdown of Fermi liquid theory which assumes that
$1/\tau\ll\omega$. Second, while the variation of scattering is
linear with frequency in both materials, in the cuprates the
curves extrapolate back to a finite positive intercept on the
$1/\tau$ axis that scales as $T$ and goes to zero at $T=0$. This
is characteristic of quantum fluctuations.\cite{varma89} For the
borocarbides the intercept at $T=0$ is on the negative $1/\tau$
side resulting in a finite intercept on the frequency
axis, which is of the order of the Debye frequency, as expected
for a Bloch-Gr\"uneisen conductivity. The Debye frequency of
LuNi$_2$B$_2$C is 250 \cm\ from a Debye temperature of 360 K
measured by neutron spectroscopy\cite{gompf97} in good agreement
with the point of intersection of the low temperature reflectance
curve with the $R=1$ axis.

A  fit of a theoretical model directly to the reflectivity avoids
the propagation of errors associated with Kramers-Kronig analysis
and the extraction of the parameters of the extended Drude model.
This is particularly relevant in a good metallic system where the
reflectivity is close to unity. The procedure is to start with a
model $\alpha_{tr}^2(\Omega)F(\Omega)$ of the electron phonon
spectral function\cite{allen71,shulga91} weighted by the amplitude
for large angle scattering and then calculate the optical
properties and compare them to experiments.

To calculate the reflectance spectra we use the extended Drude
formalism with the scattering rate given by Shulga et
al.\cite{shulga91}:
\begin{widetext}
$$
{1 \over \tau}(\omega,T)={\pi \over \omega} \int_0^{\infty}
d\Omega \alpha_{tr}^2(\Omega)F(\Omega)[2{\omega}{\coth}({\Omega
\over {2T}}) - (\omega+\Omega){\coth}({{\omega+\Omega} \over
{2T}})+ (\omega-\Omega){\coth}({{\omega-\Omega} \over {2T}})],
$$
\end{widetext}
where $T$ is the temperature, measured in frequency units.
Following Gonnelli \textit{et al.}\cite{gonnelli00} we place a restriction
on the spectral functions demanding that they fit the slope and
absolute magnitude of the dc resistivity of LuNi$_2$B$_2$C. We use
a plasma frequency of 3.3 eV in our models.

The simplest model is a single Einstein oscillator. We find that
an oscillator at 170 \cm\ with $\lambda_{tr}=0.33$ gives a fairly
good value of the room temperature resistivity and the temperature
derivative of the resistivity (35 $\mu\Omega$cm and 0.135
$\mu\Omega$cm/K, vs. the experimental values of 36 $\mu\Omega$cm
and 0.127 $\mu\Omega$cm/K). The bold curves in Fig. \ref{fig9} are the
calculated spectra based on this model.

Another model uses the phonon density of states as determined by
neutron scattering. The bare spectrum gives a rather poor fit to
the dc transport and Gonnelli \textit{et al.} found that they had to
enhance the low frequency modes (below 38 meV) by 0.7 and the high
frequency ones by 0.25, a ratio of 2.8 in the enhancement of the
low frequency spectrum. They used a $\lambda_{tr}$ of 0.53 based
on a plasma frequency of 4.25 from the Widder
\textit{et al.}\cite{widder95} optical measurements.

Because we find a lower plasma frequency of 3.3 eV we find that
somewhat different parameters than those used by Gonnelli \textit{et al.}
are necessary to fit the resistivity. Taking the Gonnelli
spectrum as a reference, we have multiplied the low frequency part
by 0.23, the upper band above 38 meV by 0.48, a ratio of 2.1 in
the enhancement of the low frequency spectrum. These parameters
were adjusted to give reasonable agreement with the temperature
dependence of the dc resistivity. The resulting spectrum is
similar to what is shown for the Einstein oscillator in Fig. \ref{fig9}.

The fit of both models to our reflectance is good at low frequency
and well within experimental uncertainties, but there are serious
deviations at high frequency which become more marked with
increasing temperature. These deviations exceed our experimental
uncertainty.

It is not possible to remove the high frequency discrepancy by
manipulating the coupling to the various phonon frequencies.  To
obtain the strong scattering seen at high frequencies it is
necessary to add oscillators above the phonon spectrum with a
continuous distribution of frequencies, a spectrum of the Marginal
Fermi liquid type as seen in the cuprate
superconductors.\cite{varma89} However such a model cannot be
justified in this case since the scattering rate curves shown
in Fig. \ref{fig8} intersect above unity, signifying the coupling to finite
frequency bosons and not to quantum critical fluctuations where
the intercept occurs at unit reflectance at low
temperature.\cite{varma89}

We think a more likely scenario is one where the threshold of
absorption seen in Fig. \ref{fig8} is a manifestation of the onset of
either multi-phonon processes or interband transitions as shown
by the 0.15 eV band in the conductivity spectra, the same
phenomenon that was responsible for the mid-infrared oscillator
shown in Fig. \ref{fig7}. There is also the possibility of a
contribution to the mid-infrared absorption by the Ni$_2$B droplets.

\section{Discussion}

We first address the question of the plasma frequency and its
variation with doping. Based on several methods of analysis we
find a Drude weight at room temperature of 3.3 eV for the undoped
sample. Other investigators have reported somewhat higher values
as a result of the inclusion of some midinfrared spectral weight
in the Drude peak giving a value closer to 4.25 eV.\cite{widder95,bommeli97}
Pickett and Singh\cite{pickett94}
calculate a plasma frequency from band structure to be 5.1 eV.
However Michor \textit{et al.}\cite{michor95} suggest, from an
analysis of specific heat data, that this is an over estimate.

\begin{figure}
 \resizebox{!}{7cm}{\includegraphics{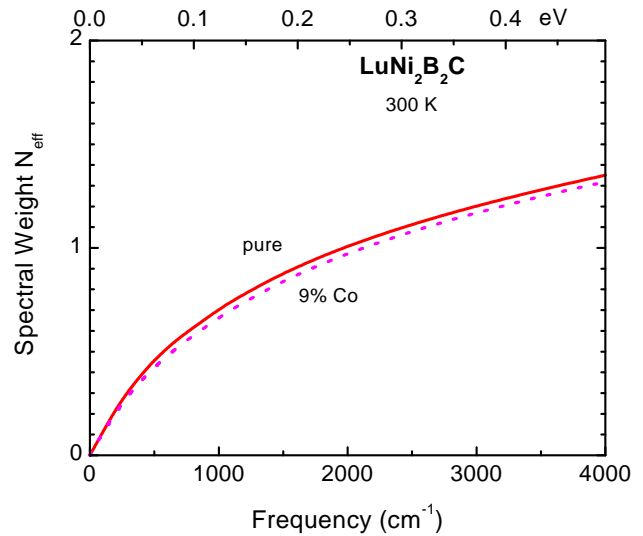}}
 \caption{\label{fig10}Partial conductivity sum rule for the pure and the doped samples.
 The small enhancement of conductivity in the pure sample would imply a small
 enhancement of the density of states at the Fermi surface. The small difference is
 not significant within our margin of error.}
\end{figure}

The question of the doping dependence of the Drude weight is more
difficult. It is clear from most of our measurements that the
changes in the spectra with Co doping are relatively slight and
the methods used in the previous paragraphs  are subject to errors
of the order of 10\% and not accurate enough for this task. To
help reduce systematic errors we use the integrated spectral
weight as a measure of Drude weight. Fig. \ref{fig10} shows the quantity
$N_{eff}$ which is the number of electrons contributing to the
conductivity in  the unit cell of volume $V_{cell}$:
$$
N_{eff}(\omega)={2mV_{cell} \over \pi e^2}\int_0^{\omega}
\sigma(\omega')d\omega'
$$
We show $N_{eff}$ at room temperature, for the pure sample, solid
curve, and the 9 \% doped sample, dashed curve.  The pure curve is
higher by about 7 \% in the region of the Drude peak. This
difference is not significant in view of our error of 0.5 \% in
reflectance which propagates to give an uncertainty of 7 \% for
the spectral weight.  The density of states at the Fermi surface
is proportional to the square of the plasma frequency and appears
to be reduced by 7 \% by the cobalt doping if we believe the
difference data of Fig. \ref{fig10}. The overall spectral weight reaches one
electron per unit cell at the frequency where the interband
contribution takes over from the Drude conductivity.

We can compare our change in Drude spectral weight with cobalt
doping  with what might be expected from the dc resistivity. From
the data of Cheon \textit{et al.}\cite{cheon98} the slope of the
resistivity changes from 0.127 to 0.140 $\mu\Omega$cm/K
in going from the pure sample to the 9 \% cobalt
doped one.  If this effect is from the change in the Drude plasma
frequency it would correspond to a change in $\omega_p^2$ of 10
\%, somewhat higher than the 7 \% that we get from the change in
the infrared Drude weight. However the dc measurements are subject
to an error of $\pm 10 $ \% due to uncertainties in contact
geometry and the difference is not significant.  It should be
noted that the dc resistivity slope is also proportional to the
coupling constant $\lambda_{tr}$ which could change with doping.
It appears that neither method, the infrared or the dc transport
is able to put an accurate limit on the change in Drude weight
with cobalt doping. All that can be said is that it appears to be
less than 10 \% from both measurements.


Our value of $2\Delta/k_BT_c=3.9 \pm 0.4$ for the doped sample
places it in the regime of weak to moderately strong coupling
superconductivity. Raman scattering on a pure sample of this
material gives a strong coupling gap ratio of 4.1,\cite{yang99} as
do thermodynamic data, such as the magnitude of the specific heat
jump.\cite{michor95} However, tunneling spectroscopy point to a
weak coupling value of the ratio.\cite{ekino96} A weak coupling
value of the gap parameter is also consistent with our electron
phonon coupling parameter $\lambda_{tr}=0.33$.\cite{carbotte90}


In summary, we have presented reflectance data on as-grown
surfaces of LuNi$_2$B$_2$C which differ in several aspects from
those measured by previous investigators along with an analysis
based on free electron theory. First our measurements confirm that
on the whole this material has the electrodynamic properties of a
good metal. We are able to fit the reflectance data at low
frequency to models of electron phonon interaction with parameters
derived from dc transport on crystals from the same source.  At
higher frequency additional absorption is observed which we
attribute to multi-phonon processes or interband transitions.  We
find a Drude plasma frequency of 3.3 eV and an electron-phonon
coupling constant $\lambda_{tr}=0.33$. This low value of
$\lambda_{tr}$ is consistent with our superconducting gap ratio of
$2\Delta/k_BT_c=3.9$ which we measure from reflectance in the
superconducting state in the cobalt doped sample which is in the
dirty limit.  We do not see any change to the plasma frequency
with cobalt doping to a level of $\pm$ 10 \% but do observe an
amount of increased scattering which is consistent with
Mattheissen's rule.

There remain several open questions. First, what is the nature of
the mid-infrared absorption band at 0.15 eV? Unlike the cuprates
where a large portion of the absorption is associated with a
Marginal Fermi Liquid spectrum of electronic coupling to the
charge carriers with an intercept at zero frequency, the intercept
here is Bloch-Gr\"uneisen-like at finite frequency. Second, the
accuracy of our experiments was insufficient to yield the detail
in scattering rate spectrum necessary to reveal the phonon
spectrum. With more accurate infrared data the $\alpha^2F(\Omega)$
spectrum of the electron phonon interaction can be mapped out as
was done for lead.\cite{farnworth76} Also, it would be important
to make accurate resistivity measurements at high temperature to
look for evidence of coupling to high frequency modes. To see
evidence for phonon effects by infrared spectroscopy, the current
single bounce reflection measurements do not have enough
resolution. Cavity measurements on large enough single crystals at
low temperature and high frequency may have enough sensitivity for
this. Also, a detailed strong coupling calculation may be able to
throw light on the possibility that the 0.15 eV band may be due to
strong coupling multi phonon processes.


The work at McMaster University was supported by the Canadian
National Science and Engineering Research Council. The Ames
Laboratory is operated for the U.S. Department of Energy by Iowa
State University under Contract No. W-7405-Eng-82. We would like
to thank J.P.~Carbotte, Hanhee Paik, and E. Schachinger for
valuable discussions.

\end{document}